\documentclass[twocolumn,aps,prl,amssymb,showpacs]{revtex4}
\usepackage{graphicx}
\usepackage{dcolumn}
\usepackage{amsmath}
\usepackage{float}

\begin{document}

\title{Coulomb cluster explosion boosted by a quasi-$dc$ pulse -- 

diagnostic tool and ultimate test of laser fusion efficiency in clusters }

\author{A. E. Kaplan }
\email{alexander.kaplan@jhu.edu}
\affiliation{Electr. and Comp. Eng. Dept, The Johns Hopkins University, Baltimore, MD 21218}

\date{\today}

\begin{abstract}
To greatly enhance output of nuclear fusion
produced neutrons in a laser-initiated Coulomb explosion of
Deuterium clusters, we propose to accelerate
the resulting ions by a quasi-$dc$ electrical pulse to the energies
where the $D^+ + D$ collision cross-section is the highest.
With $D^+$ ions bombarding then a Deuterium-rich
solid-state cathode,
this allows one to solve a few problems simultaneously by
(a) completely removing electron cloud hindering
the Coulomb explosion of ionic core, 
(b) utilizing up to $100 \%$ of the cluster ions 
to collide with the high-density packed nuclei, and 
(c) reaching highly increased cross-section of neutron production
in a single $D^+ + D$ collision,
in particular by using a multi-layered target.
We also consider the use of $E$-pulse
acceleration for diagnostic purposes.
\end{abstract}


\maketitle

Nuclear fusion reactions in solid-deuterium 
laser-produced plasma were first observed almost 50 years ago [1].
A promising recent development in the field is
generation of neutrons in laser-induced ionization
and explosion of deuterium clusters [2] with a
conversion efficiency up to $\sim 2 \times 10^6$ neutrons/J [3]. 
A  major mechanism here is 
a Coulomb explosion, CE [2-5] of the clusters,
whereby an irradiated and highly ionized cluster 
loses its free  electrons that ideally are
almost instantly swept away by the laser,
with the ionic core torn
apart by repulsive Coulomb forces resulting in CE.
Part of the process is the formation of shock-shells 
in expanding ionic cloud predicted in [6], 
explored in detail in [7] and most recently
experimentally observed in [8].
In general, however, there could be some other mechanisms
of cluster explosion, such as e. g. 
the quasi-neutral micro-plasma and
hydrodynamic models [9], etc., more characteristic for lower
laser intensities ( $< 10^{15} W/cm^2$).
While the generation of neutrons indicating nuclear fusion
reaction have been successfully  observed [1-3],
the results are still far from the goal of decades-long quest for
an elusive efficient nuclear fusion for
energy-producing application,
whereby the energy of generated 
neutrons exceeds that of the input.

Further advance in using laser-induced explosions
of clusters is blocked by three major problems:
insufficient kinetic energy of ions 
(typically a few $KeV$, with collisional cross-section,
$\sigma_{f}$, very low), $\emph vs$ $\sim 100 KeV$
where $\sigma_{f}$ goes up by
orders of magnitude, see below;
low utilization of produced ions due to
low density of surrounding plasma
(typically $<$ $10^{18} cm^{-3}$ $\emph vs$
$\sim 10^{23} cm^{-3}$ in solid state),
hence low number of fusion collisions;
and finally, free electrons that eventually neutralize
the ion cloud and hamper Coulomb explosion.
Shock-shells in CE [6,8] may increase 
that collision rate [6] by having ions collide
in the higher-density area near original cluster,
but this enhancement is still insufficient 
to attain good efficiency.

In this Letter we propose:
(1) the way of overcoming those problems by using 
a laser-synchronized electrical pulse  
to accelerate laser-produced ions $D^+$ to 
energies sufficiently high (up to $\sim 70 - 100 KeV$) to greatly
increase the cross-section
for the neutron production, 
and smashing them against a deuterium (or tritium)-rich 
solid-state target (cathode);
aside from greatly enhancing neutron output,
this would provide an ultimate
test of laser+cluster nuclear fusion 
for energy production applications,
(2) the enhancement of the solid-state
target performance by making it as a stack of thin layers,
and (3) a diagnostic of the cluster 
and explosion structure for research purposes and
application as a neutron source.

This approach may be viewed as a cross between
laser-induced CE and basic
electrostatic generation of neutrons [12]
providing substantial cross benefits.
First, there is no need anymore to strive
for powerful laser irradiation,
in particular to remove the electron cloud;
the laser energy has to be just enough 
to attain a reasonable ionization,
not to produce high ion energies 
or blow electron cloud away.
Ionized electrons here are quickly removed 
from the expanding cloud, thus barring them
from neutralizing the ions in the cloud and 
hindering the useful effect of CE.
Similarly, laser-induced ionization may allow to use
lower E-field than in electrostatic generators.
The system also simplifies the analysis 
and diagnostics of the process:
once electrons are removed, the dynamics of remaining cloud
of positive ions is strictly due to
a repulsive Coulomb explosion,
which now may differ from an ideal CE 
only in that in each case the radial 
density distribution of the ions
may be non-uniform
depending on their initial distribution.
The latter one can thus be elicited
by using segmentation of the cathode
into a few electrically isolated sections/rings
and recording the time-dependant current 
from  each one of them, see below.

In a common arrangement, a cluster or a jet
of clusters is injected between an anode collecting
ionization produced electrons
and a cathode covered by deuterium or tritium-rich material, 
as a target for $D^+$ ions accelerated by a 
strong $E$-pulse applied to the electrodes.
The ion energy then can get
much higher than that produced by CE
and reach the optimum domain of up to $100 KeV$;
it is directly controlled by the $E$-pulse amplitude.
A solid-state target insures 
then a high probability of fusion collisions.
An $E$-pulse has to be sufficiently long
to be maintained till all the ions reach the cathode.
Its duration for e. g. $50 KeV$ voltage
and $2 cm$ electrode spacing, has to be $> 10 ns$,
while the laser pulse is typically
sub-$ps$ long; the formation time
of ion cloud is even much shorter.
For a laser intensity $\gg 10^{15} W/cm^2$,
free electrons are pulled from the core
faster than a laser cycle [6],
and then they are swept away by
the $E$-pulse and brought to the anode.

Electrode geometry may vary from
spheres to cylinders, and to cones,
while a parallel configuration provides for
the simplest arrangement, Fig.1.
%
\begin{figure} [h]
\includegraphics[width=3.2in]{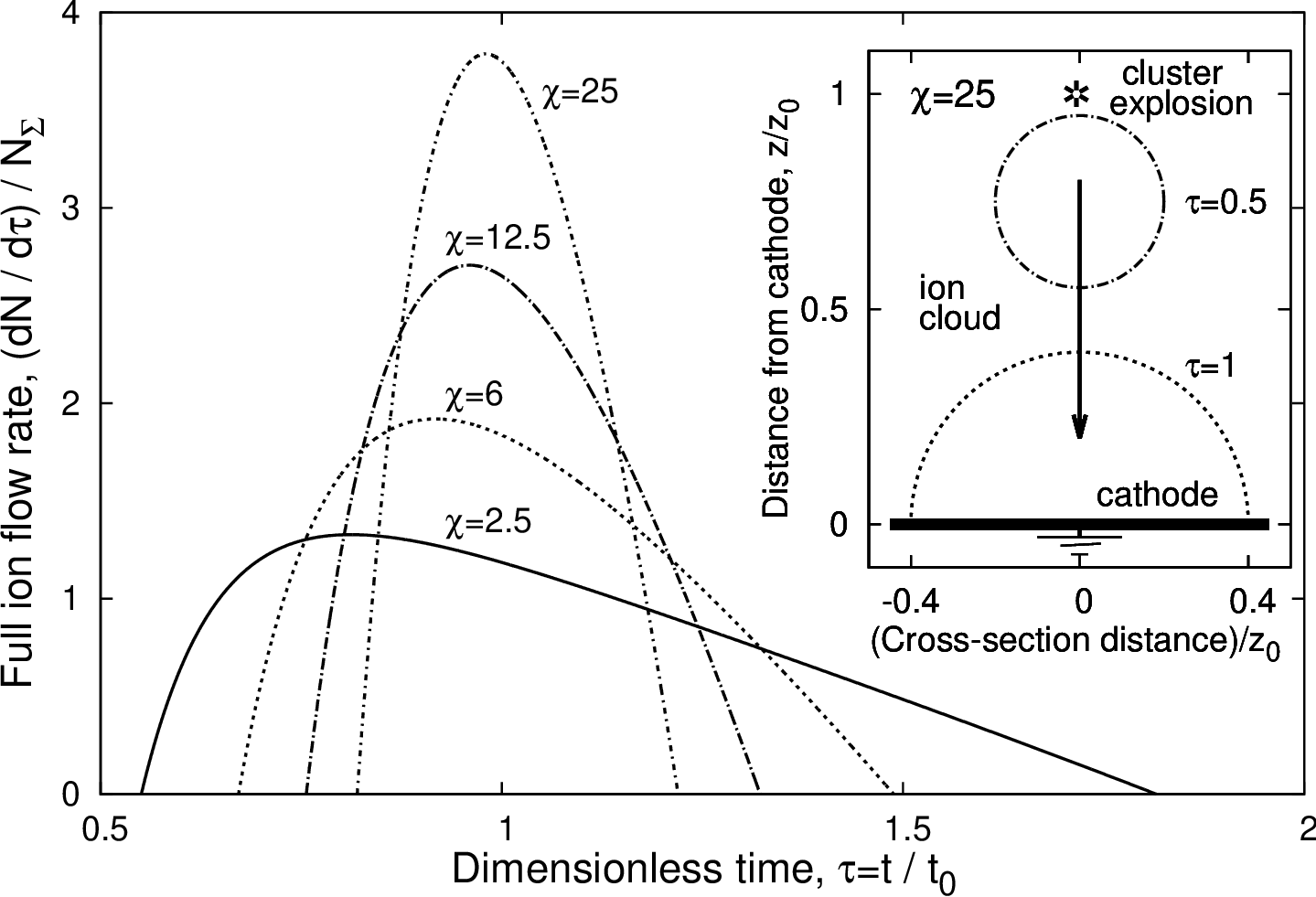}
\vspace{-0.05in}
\caption{Full ion flow rate, $ ( dN / d \tau )$ / $ N_{\Sigma} $
$vs$ dimensionless time, $\tau = t / t_0$,
for various potential/kinetic energy ratios,
$\chi = U_0 / T_{0max}$.
Inset: the schematics of the proposed experiment;
dotted circles outline
the edge of ion cloud after
cluster explosion at two different moments.
The expanding positively charged cloud moves down
toward the deuterium-rich cathode and finally
hit it under the action of high-voltage potential.}
\label{fig1}
\end{figure}
Consider a cluster
initially at the distance $z_0$ from a cathode
and a potential between them is $U_0 = q V_0$,
where $V_0$ is a respective voltage
and $q$ -- an ionic charge 
(for $D^+$, $q = e$).
Once bound electrons are freed by laser pulse,
the $E$-pulse ``vacuum-cleaning''
pulls them to the anode.
The ensuing ionic CE would then 
typically produce a shock
at the edges of a CE cloud [6-8], yet soon
the density of over-run ions decreases.
Then, for an ideal CE,
the expansion proceeds as if a sphere 
with uniform density distribution
in the momentum space.
In general, however, this
could be more complicated, due to expansion
originated by thermal explosion,
secondary ionization, 
or compositions, such as e. g. non-uniform or
heterogeneous clusters [3,13]
of different ionic species, 
or mixed clusters formed by depositing layers of
different atoms, etc.
This may result in distinctly
different initial non-uniform ion kinetic
energy radial distribution,
with the same maximum energy $T_{0max}$, see below.
In the case of an ideal CE, whereby the
initial density is almost uniform, we have [6]
\begin{equation}
T_{0max} = ( e n_i )^2 { N_\Sigma / R_0 } ,
\tag{1}
\end{equation}
where $N_\Sigma$ is a total 
number of ions in a cluster,
$n_i e$ is the ion charge 
(for hydrogen atoms or isotopes, $n_i =1$),
and $R_0$ is an original radius of cluster.
When $R_{cl} >> R_0$, where $R_{cl} (t)$
is a cloud radius, the ion  motion
is unaffected by ion collisions,
and the ion movement is inertial
in the frame of the center of mass (COM),
while COM accelerates toward a cathode with 
its $z$-speed being
$v_{_{COM}} (t) =  - U_0 t / z_0 M$,
with $t = 0$ at the moment of explosion.
Similarly to the Hubble expansion,
the ion radial distance $\rho$
from COM is proportional to their original
velocity, $\rho = v_0 t$, $v_0 \leq v_{0max}
= \sqrt{ 2 T_{0max} / M}$.

By far, the most crucial 
``boost-factor'' in the proposed scheme 
is the ion acceleration due to quasi-$dc$ potential.
We consider the simplest
neutron-producing reaction
due to collision of two deuterium nuclei:
$d + d \rightarrow {^{3}He}(0.82 MeV) + n(2.45 MeV)$.
The importance of sufficiently high nuclei energy 
transpires from the fact that the fusion cross-section 
$\sigma_{f} ( \epsilon_{in} )$,
increases with $\epsilon_{in}$
{\em {by many orders of magnitude}}.
Here $\epsilon_{in} = U_0 + T_0$ is a full 
energy of the collision of fast ion
$D^+$ with $D$ target molecules  
(such as e. g. $D_2$,
$C D_4 $, or $D_2 O$ [10,11];
$D_2$ is the most efficient one [14]).
Following [11] and
using fusion probability defined as 
\begin{equation}
w_{f}( \epsilon_{in} ) =
\int_0^{\epsilon_{in}} N_{tar} v 
\sigma_{f} ( \epsilon ) d \epsilon
/ ( d  \epsilon / dt ) , 
\tag{2}
\end{equation}
where $d \epsilon / dt$ is the energy loss rate
largely due to nucleus-nucleus scattering
(the contribution of neutron generation 
to this rate could be ignored),
$N_{tar}$ is the number density of target nuclei,
and $v$ is the velocity of deuterons.
In the range of energies of interest and 
regular solid-state density,
$w_{f}$ is approximated by a simple formula [11]
\begin{equation}
w_{f} ( \epsilon_{in} ) \approx C_D exp \left( - 
\sqrt { \epsilon_{sc} / \epsilon_{in}   }  \right)
\tag{3}
\end{equation}
where $C_D = 0.18$ and $\epsilon_{sc} = 7 MeV$.
Eq. (3) is best fitted for $D_2$ molecules;
considering an example of cluster with $N_{\Sigma} \sim 3,250$,
and $R_0 = a ( 3 N_{\Sigma} / 4 \pi )^{1/3}$
with $a \sim 10^{-8} cm$ being an initial
averaged spacing between ions
(assuming an ideal case whereby all the atoms ionized by laser),
we have $T_{0max} \sim 5 KeV$, Eq. (1).
In such a case, if one chooses
$U_0 \sim 100 KeV$,
$w_{f}$ is boosted by $\sim 0.5 \times 10^{13}$.
This makes it clear how far a Coulomb explosion 
is from producing a substantial neutron output.
However, as impressive as this number may look,
the probability $w_{f}$ per se is 
not the bottom-line to strive for,
as far as energy nuclear fusion is concerned.
For that, the minimum goal is to have
an averaged output energy of neutrons,
$\epsilon_D w_{f}$, to  
exceed that used to ionize deuterium
and accelerate $D^+$ ions, $\epsilon_{in}$,
so that the ``in/out'' efficiency $\eta$ exceeds unity,
$\eta = \epsilon_D w_{f} / \epsilon_{in} > 1$,
where $\epsilon_D = 2.45 MeV$ 
is the energy of a neutron 
exiting fusion reaction.
For real applications the input power
``at wall-plug'' must be higher, but $\eta > 1$
provides a minimal requirement.
The efficiency $\eta$ $vs$ input energy $\epsilon_{in}$
is depicted in Fig. 2;
one can see that $\eta$ never reaches unity;
it peaks at $\epsilon_{in} = \epsilon_{sc} / 4 = 1.75 MeV$
[i. e. actually outside the domain of validity of Eq. (3)],
and even then
$\eta_{max} = 4 ( \epsilon_D C_D / \epsilon_{sc} ) e^{-2} \approx 0.033 \ll 1.$
For $\epsilon_{sc} < 200 KeV$, we have $\eta < 0.006$,
and for $\epsilon_{sc} < 100 KeV$ $\eta < 10^{-3}$.
Thus, the efficient fusion is unreachable here;
essentially, this is true for any beam-target fusion,
including laser-accelerated ion beams [15].
\begin{figure} [h]
\includegraphics[width=3in]{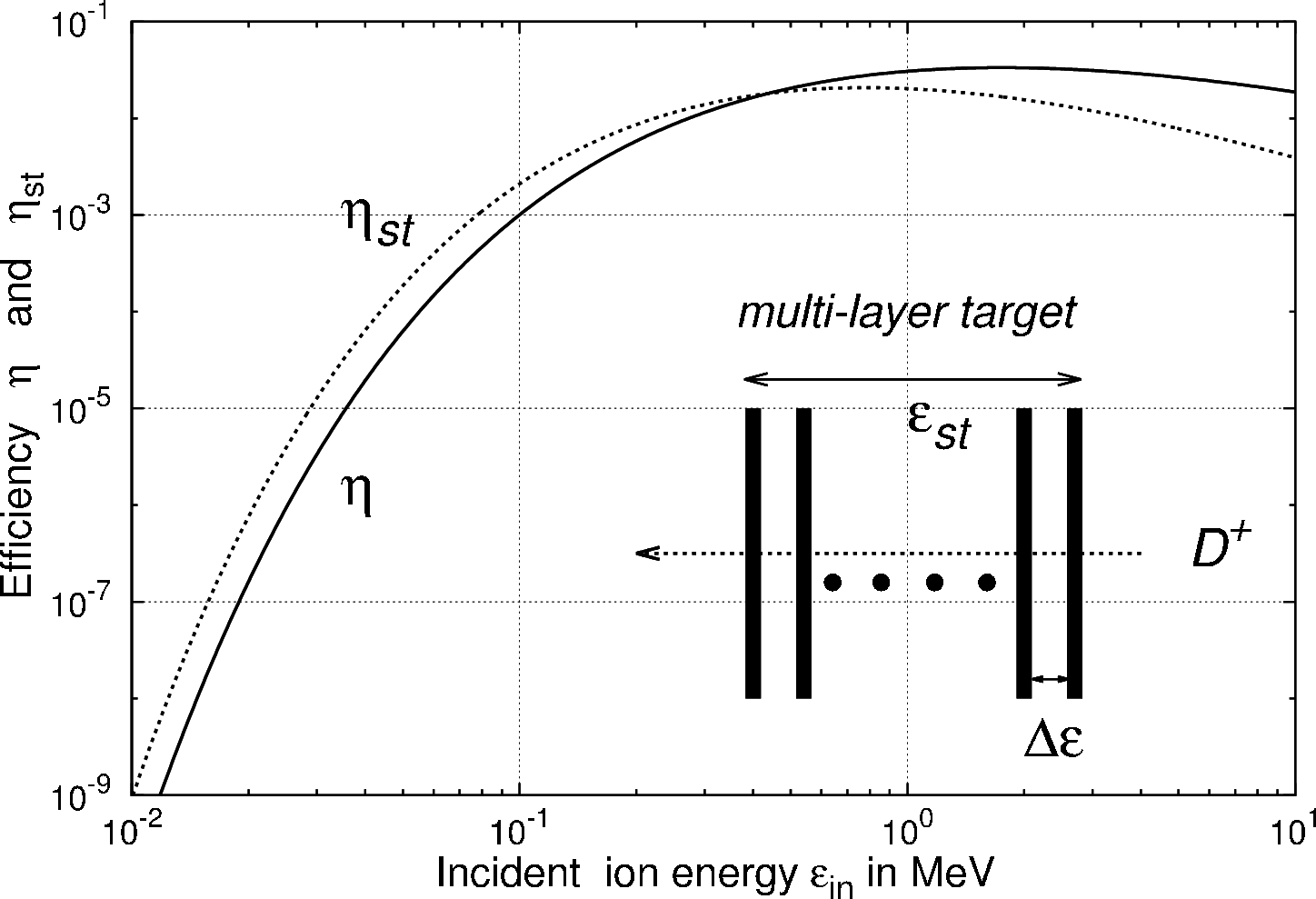}
\caption
{
The ``in/out'' efficiency of neutron production 
for bulk, $\eta$, and stack (multi-layered), $\eta_{st}$,
targets $vs$ incident energy of deuteron nucleus. 
Inset: a schematic configuration of
the stack target; $\epsilon_{_{st}}$ is 
the total potential across the stack. 
}
\label{fig2}
\end{figure}

Barring the target compression (as in inertial confinement fusion,
with a required number density up to $N \sim 10^{25} cm^{-3}$ [10]),
one way to enhance the target performance is to have it
as a stack of thin multi-layers to be crossed by ion beam,
and apply a moderate $dc$ voltage between them
to replenish the energy lost by ions.
This proposed ``cascading'' or 
``recycling'' is to mitigate the rapid ion
energy loss in a bulk target.
This loss results in dramatic reduction of
fusion cross-section $\sigma_{f} ( \epsilon )$
in Eq. (3), so that a good part of the bulk's volume
is lost for the neutron production.
Making the layers thin enough to have a low energy loss 
$\Delta \epsilon \ll \epsilon_{in}$ per layer
(to be replenished by an equal inter-layer potential), 
and their number respectively large,
$N_{st} = \epsilon_{st} / \Delta \epsilon \gg 1$,
where $\epsilon_{_{st}}$ is the total potential
across the stack,
we have the probability of fusion per layer
as $( d w_f / d \epsilon_{in} ) \Delta \epsilon$,
and the total cascade probability
$w_{st} \approx \epsilon_{st} ( d w_f / d \epsilon_{in} )$,
such that 
\begin{equation}
\eta_{st} = \frac {\epsilon_{_{D}} w_{st}} 
{\epsilon_{st} + \epsilon_{in} }
=
\eta ( \epsilon_{in} ) \sqrt { \frac  {\epsilon_{sc}} {\epsilon_{in}} } 
\frac {\epsilon_{st} / 2 } {\epsilon_{st} + \epsilon_{in} }
\tag{4}
\end{equation}
If we choose e. g. $\epsilon_{st} = \epsilon_{in}$, 
the cascade target in the realistic setting offers the
enhancement of $\sqrt { \epsilon_{sc} / \epsilon_{in} } / 4$.
over a bulk target.
In that case, $\eta_{st} ( \epsilon_{in} )$  peaks 
at $\epsilon_{in} = \epsilon_{sc} / 9 \approx 778 KeV$,
with $(\eta_{st})_{max} = 0.023$;
$\eta_{st} = 0.0021$ at $\epsilon_{in} = 100 KeV$,
and $0.0086$ at $\epsilon_{in} = 200 KeV$.
In spite of enhancement, we still have $\eta_{st} < 1$,
i. e. the efficient fusion is still unreachable.
Similar calculations for a $D^+$ ions and $Tritium$ target,
(with even higher output neutron energy, $\sim 14.1 MeV$,
instead of $2.45 Mev$ for $D^+  + D$)
leave this conclusion stand.
As was mentioned,
the same is true for any beam+target fusion mode, as
opposed to the inertial confinement fusion,
whereby the efficiency $\eta$ can in principle exceed unity.
The major reason for this
is that the product of density and confinement time
in the former case
is far below the one required 
by the so called Lawson criterion [10,11];
there is simply not enough time
for an $D^+$ ion to enter into a 
$n$-producing reaction in a bulk
target [16], while in a cascade target
it needs energy replenishment;
the situation here is remotely reminiscent
of a critical mass phenomenon in fission reactions.
A potentially viable path for
energy solution could be a hybrid
approach combining ion-beam and
inertial confinement fusion,
but it is outside the scope of this Letter.
A possible way to go might be
bombarding the solid-state target
by small ionized clusters,
instead of separate ions,
or chunks of partly-decomposed 
clusters sufficiently accelerated 
by a E-pulse, which may create
longer-sustained high-temperature 
at the impact area (see also below),
which we are planning to address elsewhere.

Regardless of energy-producing prospects,
the proposed system may be  
used both as a portable source of neutrons,
with the advantage of having orders 
of magnitude better time-clocking
of the output neutron pulse over electrostatic sources [12]
due to well synchronized laser pulse trigger,
and a greatly useful research tool
for diagnostics of cluster structure and explosion cloud.
While the diagnostics of ultra-short pulse laser-produced 
``macro''-plasma have been extensively studied (see e.g. [17]),
the nanoscale mapping  of $fs$ pulse absorption
was studied [18] recently using a "plasma explosion imaging".
Yet combined with $E$-pulse,
the system may definitely offer new possibilities.
Its diagnostics capability comes naturally
from the fact that it has already the major components 
(CE + E-field)
employed by a so called Coulomb Explosion Imaging [19-22] (CEI),
used first in 1989 [19] to yield
images of small individual molecules.
The CEI technique has been developed 
into a fine and sophisticated tool [20],
including most recent use of pixel-imaging 
mass-spectrometer camera [21],
and attaining the first image of 
the Efimov trimer in helium [22].
However, there is an important difference
with cluster CE: a cluster is comprised
of too many atoms/ions, and aside perhaps from studying
the details of its surface,
which might be of interest to 
the physics of cluster formation,
one is dealing with much more
``macro-effects'' than those in the small 
individual molecules imaging,
and might be less concerned about imaging.
As an example, an important  subject to explore
is the cluster composition, to be elicited from
analyzing the time dynamics of total ionic
current at the cathode, Fig. 1,
as well as "differential"
currents flowing through sub-cathodes, 
see below.
The advantage also
is that usually a single laser shot and 
single cluster as e. g. in [8] and [18] can be used
instead of averaging over many shots as in CEI.

To have an idea of expected 
characteristics for those applications,
we consider some details of ion flow dynamics.
We introduce a potential/kinetic energy ratio, 
$\chi = U_0 / T_{0max}$, and a dimensionless time,
$\tau = t / t_0$,
where $t_0 = z_0 \sqrt {2 M / U_0}$ is a time 
for COM to reach a cathode due to potential $U_0$ alone.
The first ions hit the cathode
in time $\tau_{min}$ and the last
ones -- in $\tau_{max}$, where
$\tau_{min,max} = \sqrt{1 + \chi^{-1}} \mp \Delta \tau / 2 ;
\Delta \tau = 2 / \sqrt{\chi}$
and $\Delta \tau$ is total duration of ion flow.
Ions with the same starting energy
$T_{0}$ make an expanding sphere 
(for its outer edge see inset, Fig. 1);
which falls down to a cathode
with acceleration $- U_0 / z_0 M$.
The rate number of ions hitting the cathode 
at $\tau_{min} < \tau < \tau_{max}$ is as:
\begin{equation}
\frac 1 {N_{\Sigma}} 
\frac { d N } {d {\tau}} 
= \frac{3 \sqrt{\chi}} {8} \left( 1 + \frac{1}{\tau^2} \right) 
F ( \tau ) ;  \ \ \
F = {\int}_{\xi_1}^{\xi_2} f ( \xi ) d \xi ,
\tag{5}
\end{equation}
where $\xi = {T_0} / {T_{0max}}$
is a relative initial kinetic energies of ions,
and function $f ( \xi )$ describes a radial 
distribution of these energies that satisfies
a condition 
$ {\int}_0^1 f ( \xi ) d ({\xi^{3/2}}) = 1 $.
In the case of ideal CE, we have $f_{CE} = const = 1$.
To illustrate density/energy-profile sensitivity of this system,
we will also consider two other distinctly
different models of that distribution,
in particular ``hot ball'', $f_{HB} = 5 \xi / 3$,
which has a hollow core, while its
outer shell is populated by hottest ions,
thus making it a sustained shock,
and a ``cool ball'', $f_{CB} = 5 ( 1 - \xi ) / 2$,
with a dense cold core and a vanishing density
of ``hot'' outer ions.

In Eq. (5), the integration
limits $\xi_1 ( \tau )$ and $\xi_2 ( \tau )$ are
determined by the area of the cathode engaged.
If all the ions hitting cathode are included,
we have $ \xi_2 = 1$, and
$\xi_1 = \xi_{min} ( \tau ) =  ( {\chi} /  4 )
\left( {\tau^{-1}} - \tau \right)^2 < 1 $,
which is a minimal initial energy of ions
reaching a cathode at the moment $\tau$, so that
for CE, $F_{CE} = 1 - \xi_{min}$,
for hot ball, $F_{HB} = 5 ( 1 - \xi_{min}^2 ) / 6$,
and for a cool ball, $F_{CB} = 5 ( 1 - \xi_{min} )^2 / 4$.
For the CE case,
the rate ${ d N} / {d \tau}$ in units $N_\Sigma$
for $\chi$ from $2.5$ to $25$ is depicted in Fig. 1
(if ${T_{0max}} = 4 KeV$, that would correspond to $U_0$
ranging from $10$ to $100 KeV$).
While initial kinetic energy of an ion is $T_0$,
which increases to $T = U_0 + T_0$ when it hits a cathode,
the total energy of the cloud delivered to the cathode
during entire process in an ideal CE case,
is $T_\Sigma = N_\Sigma ( U_0  + 3 T_{0max} / 5 )$.

To prevent ions from hitting an anode,
one needs a sufficient overhead spacing, 
$z_{up}$, between clusters and anode,
(with total cathode-anode spacing 
$z_{CA} = z_0 + z_{up} \geq z_{cr}$,
and thus a sufficient voltage
between the plates, $V_{CA} \geq V_{cr}$), 
so that 
${z_{cr}} / {z_0} = {V_{cr}} /  {V_0} = 1 + \chi^{-1}$.
The maximum ``hot spot'' radius $\rho_{sp}$ at the cathode
is reached at $\tau_{sp}$, which are respectively as:
${\rho_{sp}} / {z_0} =  2 \sqrt {1 + \chi } / \chi$,
$\tau_{sp} = \sqrt{  2 \chi^{-1} + 1 }$,
i. e. the spot gets tighter as 
$\chi$ increases, as expected.
The effect is sensitive to the 
ion energies distribution $f (\xi )$ in the 
explosion, and thus may offer well-resolved
time-of-flight diagnostics.
It can be implemented by segmenting the cathode
into isolated concentric rings or sub-cathodes,
and recording ion flow in each
of them, as well as a total count of the ions,
for various distributions $f (\xi )$, i
as illustrated in Fig. 3 for three rings,
The calculations here are based on Eq. (5),
where we consider three model radial density profiles: 
(a) an ideal Coulomb explosion, $f_{CE} = const = 1$,
(b) ``hot ball'' profile, $f_{HB} = 5 \xi / 3$,
with a hollow core, while its
outer shell is populated by hottest ions,
thus making it a sustained shock, and 
(c) a ``cool ball'' profile, $f_{CB} = 5 ( 1 - \xi ) / 2$,
with a dense cold core and a vanishing density
of ``hot'' outer ions.
For the demonstration purposes we consider
here the set of three rings, the central one
being a disc with a radius $\rho < \rho_1$,
a middle ring $\rho_1 < \rho_c <  \rho_2$,
and external ring $\rho_2 < \rho_c < \rho_{max}$,
where we set the sizes in such a way that in the end
of the process, the total ion flow in each of them
be the same for all three 
sub-cathodes for the ideal CE case,
whereby they have to be
$\rho_1 / \rho_{sp} \approx 0.475$ and 
$\rho_2 / \rho_{sp} \approx 0.709$.
\vspace{.1in} \\
Fig. 3 depicts the time dynamics of the current/flow 
of ions through each ring and total ion count (see inserts)
in each of those rings for each chosen model.
It illustrates a great potential of such a system
for spectroscopy and diagnostic purposes.
The ion flow dynamics for various models
in Fig. 1 clearly indicates that the effect is
sensitive to the distribution
of ion energies $f (\xi )$ in the original
explosion, and thus may offer a well-resolved
time-of-flight diagnostics of that distribution.
Depending on application, 
the electrodes geometry can be made as
spherical or cylindrical surfaces.
In general case, instead of rings,
the cathod can be made of multiple
pixels that would provide for much
greater temporal and spatial resolution
of explosion ion cloud and its dynamics.
\begin{figure*} 
\begin{minipage}{\textwidth}%
\leftline{\includegraphics[angle=270,width=2.35in]{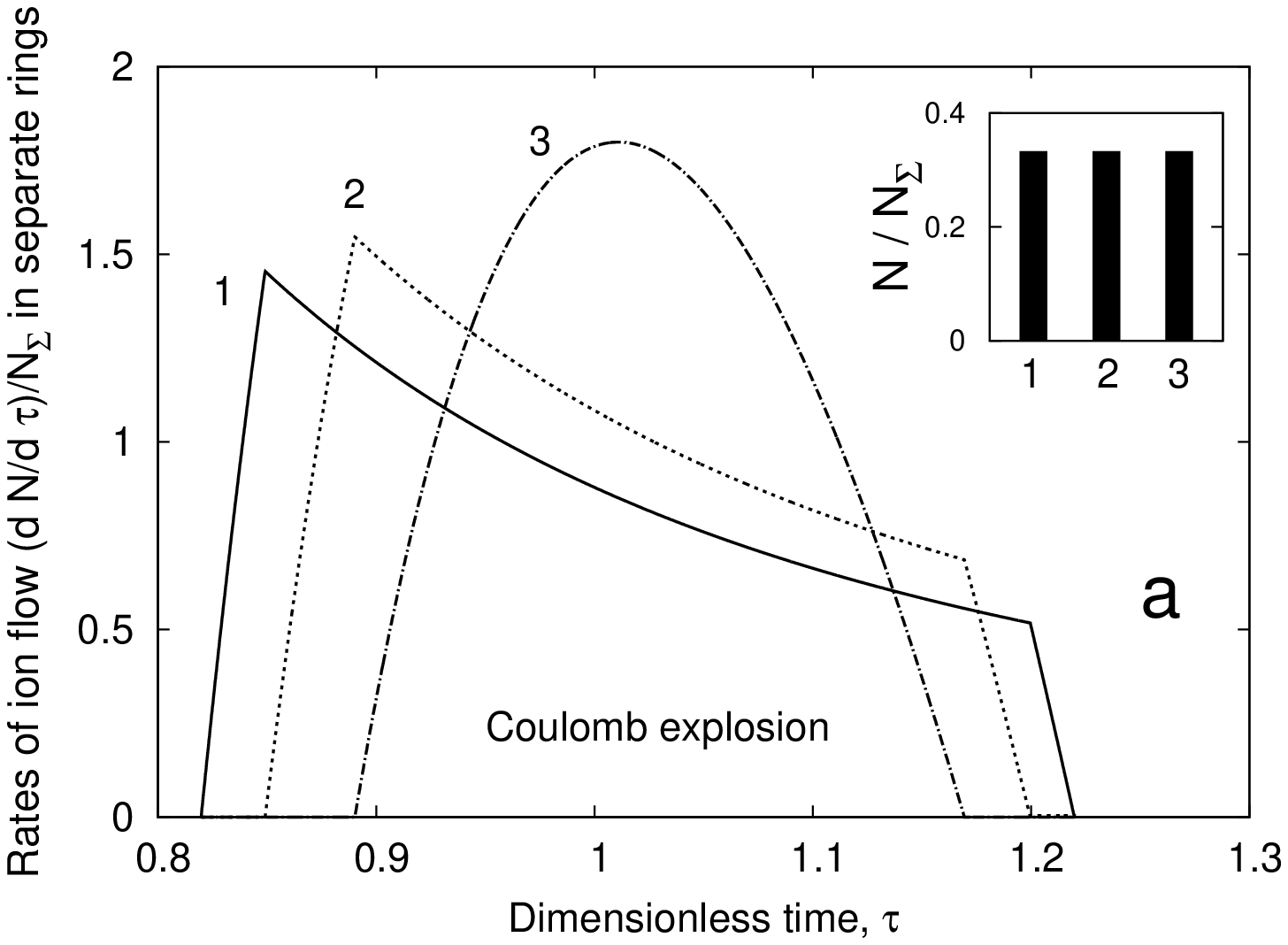}}
\end{minipage}%
\begin{minipage}{-.33\textwidth}
\rightline{\includegraphics[angle=270,width=2.35in]{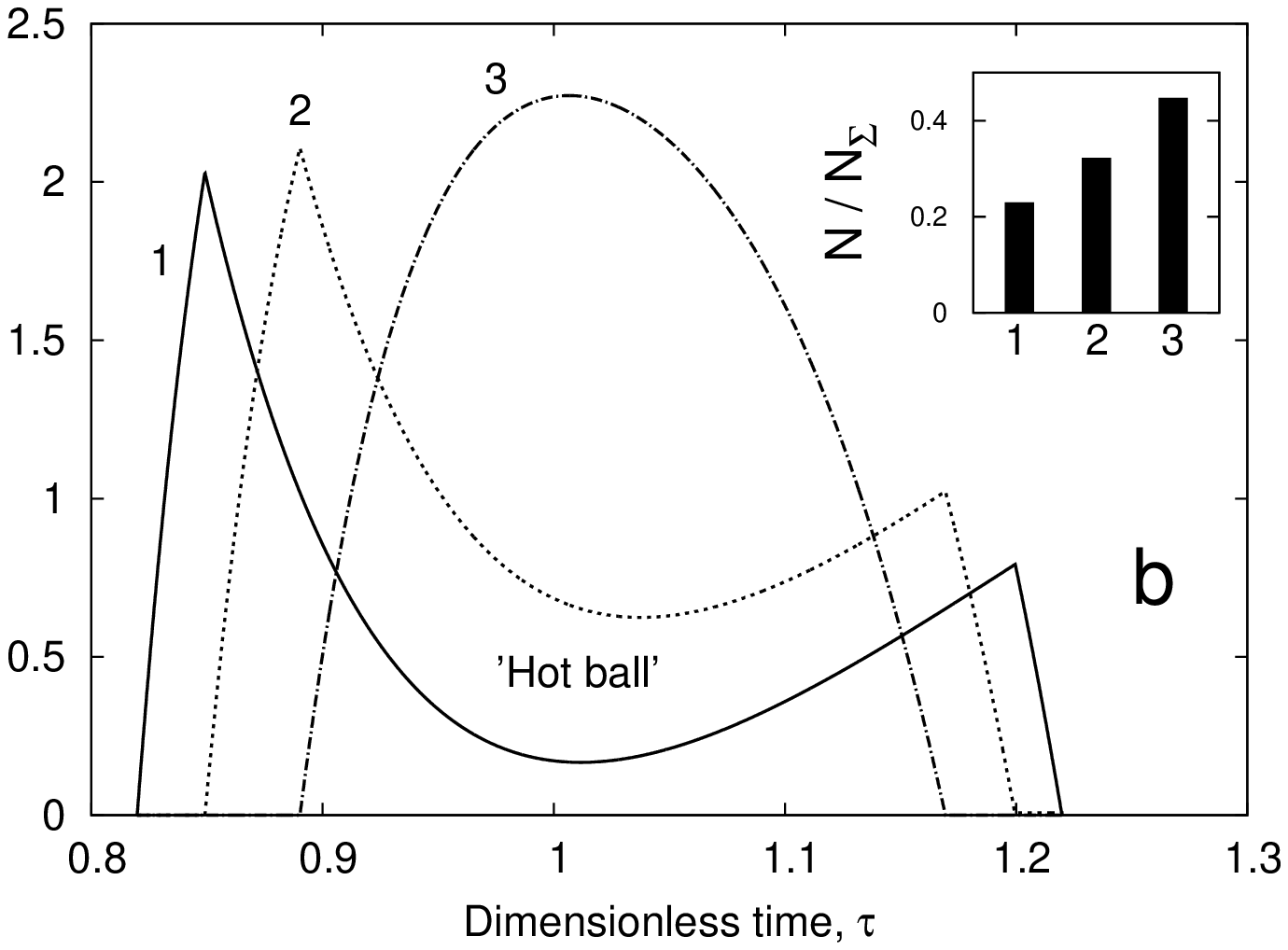}}
\end{minipage}%
\begin{minipage}{.0\textwidth}
\rightline{\includegraphics[angle=270,width=2.35in]{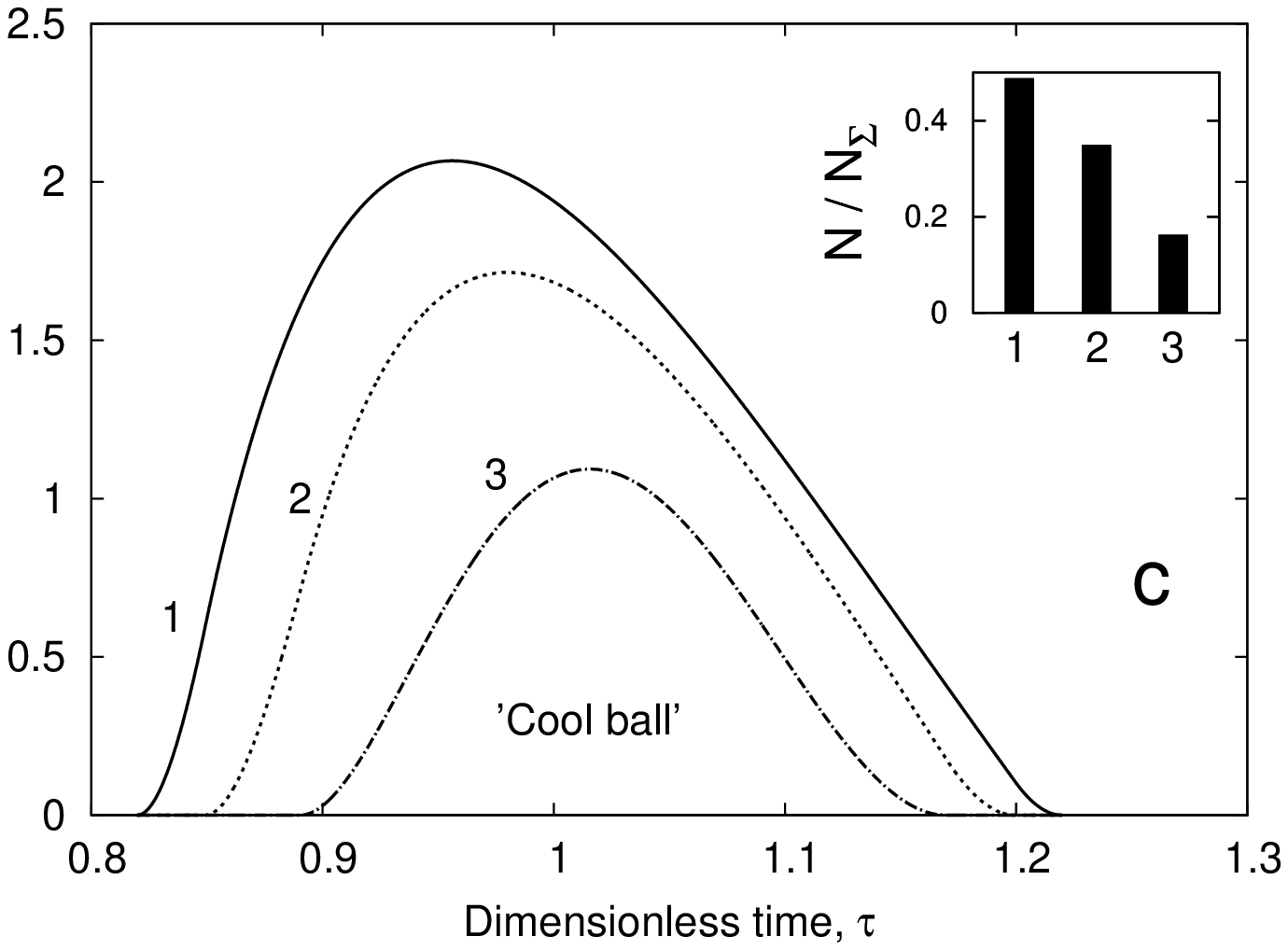}}
\end{minipage}%
\vspace{-0.1in}

\begin{minipage}{\textwidth}%
\caption 
{
Rates of the flow of ions through sub-cathode rings
(1 -- inner, 2 -- medium, 3 -- outer rings),
and time-integrated ion count (inserts)
in each of those rings for various models:
(a) an ideal Coulomb explosion (CE) $f_{CE} = const = 1$,
with almost even distribution of ions inside the cloud;
(b) a ``hot ball'', $f_{HB} = 5 \xi / 3$,
with a hollow core and outer shell of hot ions,
making a sustained shock; and
(c) a ``cool ball'', $f_{CB} = 5 ( 1 - \xi ) / 2$,
with a dense colder core and vanishing density of ``hot'' outer ions.
}
\label{fig3} %
\end{minipage} %
\end{figure*}
Ultimately, the cathode can be made
as a large set of tiny pixels
(with resolution determined then
by pixel numbers), as well as to
serve as a mass-spectrometer tool 
in search and production of multi-atomic "nano-chunks"
to be explored in the attempt
to bridge ion-beam and inertial confinement fusion.

In conclusion, we proposed to reach 
ultimately high yield of neutrons
in laser-driven explosion of deuterium clusters,
by using a synchronized $E$-pulse
with up to $100 KeV$ peak field,
and making $D^+$ ions 
bombard a negatively charged deuterium-rich cathode. 
The maximum yield is reached as
free electrons are removed from
the ion cloud, $D^+$ ions are fully utilized
and made to collide 
with solid-state deuterium-rich
target instead of plasma.
We also proposed to further enhance
the output by using a multi-layer target.
While the energy production goal
appears to be still unreachable,
the major application of the system
could be an efficient neutron source
with laser-controlled timing of neutron pulses.
The system can also be made into
a sensitive diagnostic tool to resolve 
an intrinsic structure of ion cloud and cluster itself.


\vspace{-.2in}
\begin{thebibliography}{99}
%
\bibitem{b1, firts fusion in laser explosion}
F. Floux, D. Cognard, L.-G. Denoeud, G. Piar, D. Parisot, J.
L. Bobin, F. Delobeau, and C. Fauquignon, 
Phys.  Rev. A $\bf 1$, 821 (1970).
%
\bibitem{b2, recent nuclear fusion with lasers}
T. Ditmire, J. Zweiback, V. P. Yanovsky, T. E. Cowan, G.
Hays, and K. B. Wharton, 
Nature (London) $\bf 398$, 489–492 (1999);
J. Zweiback, R. A. Smith, T. E. Cowan, G. Hays, K. B. Wharton,
V. P. Yanovsky, and T. Ditmire, 
Phys. Rev. Lett. $\bf 84$, 2634 (2000);
G. Grillon, Ph. Balcou, J.-P. Chambaret, D. Hulin, J. Martino,
S. Moustaizis, L. Notebaert, M. Pittman, Th. Pussieux,
A. Rousse, J.-Ph. Rousseau, S. Sebban, O. Sublemontier,
and M. Schmidt, 
Phys. Rev. Lett. $\bf 89$, 065005 (2002).
%
\bibitem{b3 Efficient fusion neutron generation from heteronuclear clusters}
H. Y. Lu, J. S. Liu, C. Wang, W. T. Wang, Z. L. Zhou, 
A. H. Deng, C. Q. Xia, Y. Xu, X. M. Lu, Y. H. Jiang,
Y. X. Leng, X. Y. Liang, G. Q. Ni, R. X. Li, and Z. Z. Xu,
Phys. Rev. $\bf A80$, 051201 (2009).
%
\bibitem{b4 Solid experim. confirmation of Coulomb-explosion mechanism}
Gotts, N. G., Lethbridge, P. G. and Stace, A. J.
J. Chem. Phys. 96, 408-421 (1992);
K. W. Madison, P. K. Patel, M. Allen, D. Price, T. Ditmire,
J. Opt. Soc. Am. B $\bf 20$, 113 (2003).
%
\bibitem{b5,Explosion Dynamics of Rare Gas Clusters in Strong Laser Fields}
M. Lezius, S. Dobosz, D. Normand, and M. Schmidt,
Phys. Rev. Lett. $\bf 80$, 261 (1998);
G. A. Mourou, T. Tajima, and S. V. Bulanov,
Rev. Mod. Phys., $\bf 78$, 309 (2006);
Th. Fennel, K.-H. Meiwes-Broer, J. Tiggesbäumker
P.-G. Reinhard, P. M. Dinh and E. Suraud,
$\it ibid$, $\bf 82$, 1795 (2010);
K. I. Popov, V. Yu. Bychenkov, W. Rozmus, and L. Ramunno,
Phys. Plasmas $\bf 17$, 083110 (2010).
%
\bibitem{b6} 
A. E. Kaplan, B. Y. Dubetsky, P. L. Shkolnikov,
Phys. Rev. Letts, $\bf 91$, 143401 (2003).
%
\bibitem{b7} F. Peano, R. A. Fonseca, and L. O. Silva,
Phys. Rev. Lett. $\bf 94$, 033401 (2005);
V. Yu. Bychenkov and V. F. Kovalev,
Plasma Phys. Reports, $\bf 31$, 178 (2005);
%
\bibitem{b8, single cluster shock wave, experiment} 
D. D. Hickstein, F. Dollar, J. A. Gaffney,
M. E. Foord, G. M. Petrov, B. B. Palm, K. E. Keister,
J. L. Ellis, C. Y. Ding, S. B. Libby, J. L. Jimenez,
H. C. Kapteyn, M. M.  Murnane, and W. Xiong,
Phys. Rev. Letts, $\bf 112$, 115004 (2014).
%
\bibitem{b9, quasi-neutral plasma explosion; hydrodynamic model of explosion} 
A. V. Gurevich, L. V. Pariskaya, and L. P. Pitaeveskii,
Sov. Phys. JETP, $\bf 22$, 449 (1966), also $\bf 36$, 274 (1973);
J. E. Allen and J. G. Andrews, J. Plasma Phys. $\bf 4$, 187 (1970);
L. M. Wickens, J. E. Allen, and P. T. Rumsby, 
Phys. Rev. Lett. $\bf 41$, 243 (1978);
M. A. True, J. R. Albritton, and E. A. Williams, 
Phys. Fluids $\bf 24$, 1885 (1981);
H. M. Milchberg, S. J. McNaught, and E. Parra,
Phys. Rev. $\bf E64$, 056402 (2001).
%
%
\bibitem{b10, Croos-section of neutron generation by deuterium collisions}
S. Atzeni and J. Meyer-ter-Vehn,
``The Physics of Inertial Fusion'',
Oxford University Press, 2004.
%
\bibitem{b11 Cross-section of neutron generation!!}
V. P. Krainov and B. M. Smirnov, Sov. Phys. JETP, $\bf 105$, 559 (2007).
%
\bibitem{b12, electrostatic generators of neutrons}
J. Reijonen, K.-N. Leung, R. B. Firestone, J. A. English, 
D. L. Perry, A. Smith, F. Gicquel, M. Sun,  H. Koivunoro, T.-P. Lou, 
B. Bandong, G. Garabedian, Zs. Revay, L. Szentmiklosi, G. Molnar,
Nucl. Instrum. and Meth., $\bf A522$, pp.598-602, (2004);
J. Csikai, Ed. ``Handbook of Fast Neutron Generators''; 
CRC Press: Boca Raton, FL, 1987; Vols. I and II, 242 and 224 pp.
%
\bibitem{b13}
H. Pauly, ``Atom, Molecule and Cluster Beams'' (Springer,
Berlin, 2000), Vol. II, Sec. 2.6.
%
\bibitem{b14}
H. S. Bosch and G. M. Hale, Nucl. Fusion, $\bf 32$, 611 (1992).
%
\bibitem{b15 Laser accelerated ions in ICF research prospects and experiments}
For a review of ion acceleration by ultra-intense lasers, see e. g.
M. Roth, E. Brambrink, P. Audebert, M. Basko, A. Blazevic, R. Clarke, J. Cobble,
T. E. Cowan, J. Fernandez, J. Fuchs, M. Hegelich, K. Ledingham,
B. G. Logan, D. Neely, H. Ruh, and M. Schollmeier,
Plasma Physics and Controlled Fusion {\bf 47}, B841 (2005)
%
\bibitem{b16 Low probabilty of neutron generation!!}
Be reminded that a neutron generation
reaction requires a collision by two deuterium nuclei,
which due to huge repulsing Coulomb potential
is much less probable then
atom-ion collisions between $D$ and $D^+$.
%
\bibitem{b17 diagnostics of laser-produced plasma}
M. Roth, J. Instrumentation, {\bf 6}, R09001 (2011).
ICAL REVIEW A 91, 053424 (2015)

\bibitem{b18 laser-diagnostics of CE}
D. D. Hickstein, F. Dollar, J. L. Ellis, K. J. Schnitzenbaumer, 
K. E. Keister, G. M. Petrov, C. Ding, B. B. Palm, J. A. Gaffney, 
M. E. Foord, S. B. Libby, G. Dukovic, J. L. Jimenez, 
H. C. Kapteyn, M. M. Murnane, and W. Xiong,
ACS Nano, {\bf 8}, 8810 (2014).
%
\bibitem{b19 FIRST CE imaging}
Z. Vager, R. Naaman, E. P. Kanter,
Science, {\bf 244}, 424 (1989).
%
\bibitem{b20 suggested by the Ref. b}
H. Hasegawa, Chem.Phys.Lett. {\bf 349}, 57 (2001);
C. Cornaggia, Laser Physics, {\bf 19}, 1660 (2009);
H. Xu, T. Okino, K. Nakai, and K. Yamanouchi
in "Progress in Ultrafast Intense Laser Science VII",
K. Yamanouchi, D. Charalambidis, D. Normand, Eds,
Springer Series in Chemical Physics {\bf 100}, 35 (Heidelberg, 2011);
%
\bibitem{b21 suggested by the Ref. b;  pixel-imaging maas-spectrometer camera}
C.S. Slater, S. Blake, M. Brouard, A. Lauer,
C. Vallance, C. S. Bohum, L. Cristensen, J. Nielsen,
M. P. Johansson, H. Stapelfeldt, Phys.Rev. {\bf A 91}, 053424 (2015);
%
\bibitem{b22 suggested by the Ref. b;  pixel-imaging maas-spectrometer camera}
M. Kunitski, S. Zeller, J. Voigtsberger, A. Kalinin,
L. Ph. H. Schmidt, M. Sch\"{o}ffler, A. Czasch, W. Sch\"{o}llkopf,
R. E. Grisenti, T. Jahnke, D. Blume, and R. D\"{o}rner,
Science, {\bf 348}, 551 (2015).
%
%
\bibitem{cit23}
See  Supplemental Material at [URL will be inserted by publisher]
for the demonstration example of the diagnostics of the cluster density
profile for different models of those profiles
using the set of three sub-cathodes.
%
\end{thebibliography}
\end{document}